\documentclass[prb,aps,twocolumn,amsmath,amssymb,floatfix,superscriptaddress,nobibnotes]{revtex4}
\usepackage[dvips]{graphics}
\usepackage{color}
\definecolor{dred}{rgb}{0,0,0.6}
\usepackage{graphicx}  
\usepackage{dcolumn}   
\usepackage{bm}        
\usepackage{amssymb}   
\usepackage{amsmath}
\usepackage{braket}
\usepackage{upgreek}
\usepackage[mathscr]{euscript}
\usepackage{color}
\usepackage{float}
\usepackage[toc,page]{appendix}
\usepackage[normalem]{ulem}

\DeclareUnicodeCharacter{2212}{-}

\begin{document}

\title{Graphene/hBN heterostructure based Valley transistor: Dynamic Control of valley current in synchronized nonzero voltages, within the time-dependent regime}

\author{A. Belayadi}
\email{abelayadi@usthb.dz}
\affiliation{University of Science and Technology Houari Boumediene, Bab-Ezzouar, Algeria.}

\author{C. I. Osuala }
%\email{cosuala@stevens.edu}
\affiliation{Department of Physics, Stevens Institute of Technology, Hoboken, NJ 07030 USA.}

\author{I. Assi}
%\email{iassi@mun.ca}
\affiliation{Department of Physics and Physical Oceanography,
Memorial University of Newfoundland and Labrador, St. John’s, Newfoundland \& Labrador, Canada A1B 3X7.}

\author{A. Naif}
%\email{s201626960@kfupm.edu.sa}
\affiliation{Department of Physics, National University of Singapore, Singapore}

\author{J. P. F. LeBlanc}
%\email{jleblanc@mun.ca}
\affiliation{Department of Physics and Physical Oceanography,
Memorial University of Newfoundland and Labrador, St. John’s, Newfoundland \& Labrador, Canada A1B 3X7.}
\affiliation{Compute Everything Technologies Ltd., St. John's, Newfoundland \& Labrador, Canada.}

\author{A. Abbout}
\email{adel.abbout@kfupm.edu.sa }
\affiliation{Department of Physics, King Fahd University of Petroleum and Minerals, 31261 Dhahran, Saudi Arabia.}
\affiliation{Quantum Computing Group, Intelligent Secure System (ISS) center,
King Fahd University of Petroleum and Minerals, 31261 Dhahran, Saudi Arabia}

%%----------------------------------
\begin{abstract}

Graphene/hexagonal boron nitride (hBN) heterostructures represent a promising class of metal–insulator–semiconductor systems widely explored for multifunctional digital device applications. In this work, we demonstrate that graphene, when influenced by carrier-dependent trapping in the hBN spacer, triggered by a localized potential from Kelvin probe force microscopy (KPFM), can exhibit the behavior of a valley transistor under specific conditions.
We employ a tight-binding model that self-consistently incorporates a Gaussian-shaped potential to represent the effect of the tip gate. Crucially, we show that the heterostructure can function as a field-effect transistor (FET), with its operation governed by the bias gate (which shifts the Fermi level) and the tip-induced potential (which breaks electron–hole symmetry by selectively trapping electron or hole quasiparticles).

Our results reveal that, under specific conditions involving lattice geometry, pulse frequency, and gates voltages, the device exhibits valley transistor functionality. The valley current ($I_{K_1=−K}$ or $I_{K_2=+K}$) can be selectively controlled by synchronizing the frequencies and polarities of the tip and bias gate voltages. Notably, when both gates are driven with the same polarity, the graphene channel outputs a periodically modulated, pure valley-polarized current. This enables the device to switch between distinct {\bf ON/OFF} valley current states even at finite bias.
Interestingly, when the $I_{K_1=−K}$ current is in the {\bf ON} (forward current) state, the $I_{K_2=+K}$  current is {\bf OFF}. Reversing the gate polarity inverts this behavior: $I_{K_1=−K}$ becomes {\bf OFF}.  while $I_{K_2=+K}$) turns {\bf ON} (reversed-current). These findings pave the way toward realizing low-voltage valley transistors within metal–insulator–semiconductor architectures, offering new avenues for multifunctional applications in valleytronics and advanced gating technologies.

\end{abstract}

\maketitle

%%----------------------------------------------------------------------------------------------------
\section{introduction}\label{sec1}
Graphene-based devices, with a two-dimensional layered van-der-Waals heterostructure, play a major role in various areas including electronics \cite{A.K.Geim, A.E.Curtin}  and spintronics \cite{T.Frank, zfghg, Juan.F.Sierra, K.Zollner-2021, T.Naimer, belaya-tmd}. Recently, it has been shown that Dirac points, denoted as $K_1=+K$ and $K_2=-K$, can serve as alternative degrees of freedom for encoding and manipulating information via valley-specific transport channels. This emerging approach defines the basis of a new research domain known as ``valleytronics'' \cite{Milovan, Rycerz, Pereira, Vozmediano, Settnes,Kumari2025}. Analogous to spintronics, where the electron’s spin is exploited for information processing  \cite{costa, lltoa, yi-wen, d-zamb, Aktor, Fu}  we use valley-polarized currents to perform similar task. Additionally, graphene/hBn heterostructure are known to be promising material, for several applications due to their high carrier mobility \cite{Young, Zomer, ZTang, uddin2025emerging, Veyrat, Sonntag} with a valence and conduction bands touching at both valleys. This provides a tunable Fermi-level \cite{Milovan} and a membrane that can be locally altered in the presence of scanning tunneling microscopy (STM) \cite{Klimov,  Wong, Juwon}.

For instance, metal-insulator-semiconductor (MIS) type devices such as graphene/hBN deposited on SiO$_2$/Si substrate have beenshown to be good candidates for ideal charge injection through the graphene interface because of the 2D quantum tunneling taking place between semiconducting channels and metal \cite{Lee, Britnell, Gaskell, Jung, Zhang, Bai}. It is therefore considered a good MIS device to incorporate in several digital technologies such as Field-effect transistors and memory devices \cite{Kaushal, Liu, Iosu, Das, Beck, Li, Tanrkulu, Mukherjee}. To this end, further advancement in graphene/hBN applications is required, as many challenges remain to be addressed. By adopting electron valleys as a new degree of freedom for information processing, it is possible to generate a valley-polarized current, potentially enabling synthetic valley transistor devices.  
Similar to conventional spintronic systems, where spin injection and detection are employed to manipulate spin currents, the realization of reliable valley injection and control could represent a significant step forward. Enhancing valley integration within such heterostructures is therefore anticipated to be a major milestone in the development of next-generation electronic and valleytronic applications.

It is known that  valley-polarized currents might take place when local strain is applied to a graphene sample due to  induced pseudo-magnetic fields (PMFs), which affect differently each valley \cite{Milovan, Settnes, Naif}. Notably, it will not be possible to induce  PMFs by straining all layers in  an MIS-type device. For this purpose, we would like to follow a method that does not involve any PMFs or real external magnetic fields (RMFs). 

In semiconducting materials, the quasi-bound states (QBS), at lower energy (electrons),  populate the valence bands while being  separated by an energy gap from quasi-bound states, at higher energy  (holes), in the conduction bands. In some materials, the transitions of the QBS from the valence to the conduction band across the band gap are processed by the same energy with a momentum direction related to the orientation of the axes in the reciprocal space. This momentum dependence is associated with the valleys, where electrons occupy one valley while holes populate states in the other valley. This breaking of electron-hole symmetry results in lifting the valley degeneracy. 

Fortunately, many studies have developed alternative methods to tune the potential symmetry while preserving time-reversal symmetry (magnetic properties), without altering the electronic phase transitions, and ensuring unhindered current flow  at perpendicular incidence \cite{Rycerz, Gunlycke1, YLiu, Freitag, Samaddar, Fehske}. In this context, graphene/hBN heterostructure is adopted as the main MIS device, where several studies have shown that a radial perturbative potential represented by either a power function or a Gaussian might be advantageously employed to modify its band spectrum. Meanwhile, it can break the symmetry between the valence and conduction bands where the electronic properties are strongly relying on the quasi-bound states at lower energy electrons or higher energy holes (QBS) \cite{Orlof,Wyatt, SYLi, Samaddar}. More importantly, we refer to scanning tunneling microscopy \cite{MFreitag} and Kelvin probe force microscopy (KPFM)  \cite{Wyatt, SYLi} which both have been  used to probe the potential landscape of the electronic structure in graphene/hBN heterostructures. Those methods have used a tip-perturbative potential and revealed the presence of broken quasi-bound states between the conduction band (electrons) and valence band (holes).

To this end, the opportunity for integrating valley-polarized current is advantageously expected in graphene/hBN heterostructure, with broken potential symmetry, since Klein tunneling sorts the induced electrostatic potentials to be transparent at normal incidence to the quasi-bound states in the valence and conduction bands \cite{Fehske, Rycerz, Cheianov}.

The main purpose of the present contribution is to propose an MIS-model device, based on graphene/hBN, as an elementary unit  for valley-field effect transistors operating at low bias and outputting {\bf ON/OFF} valley currents with a  periodic modulation with time. The proposed model is anticipated to perform effectively at low temperatures, utilizing induced electrostatic potentials as local radial gates to selectively trap carriers based solely on their valley index provided specific conditions are met. We demonstrate that, under well-defined parameters involving the magnitude and polarity of the bias voltage and the induced potential, the device can efficiently generate a valley-polarized current. In this regime, distinct {\bf ON/OFF} switching of valley currents occur independently, even under finite bias conditions, and with remarkably low power consumption. 

Our manuscript is organized as follows. In Sec. \ref{sec2}, we introduce the graphene/hBN MIS-like device in the presence of a perturbative potential landscape induced, for instance, by KPFM. We also describe the adopted tight-binding model that computes and maps the current flow depending on each valley. In Sec. \ref{sec3}, the numerical results are presented and discussed where we mainly show how the values and signs of the bias voltage and induced potential are correlated to operate {\bf ON/OFF} valley-polarized current. Finally, a summary of the main findings and concluding remarks are provided in Sec. \ref{sec4}.
%%----------------------------------------------------------------------------------------------------
\section{Model and Methods}\label{sec2}
\begin{figure}[tp]
\centering
%\hspace*{-0.4cm}
\includegraphics[width=9cm, height=8cm]{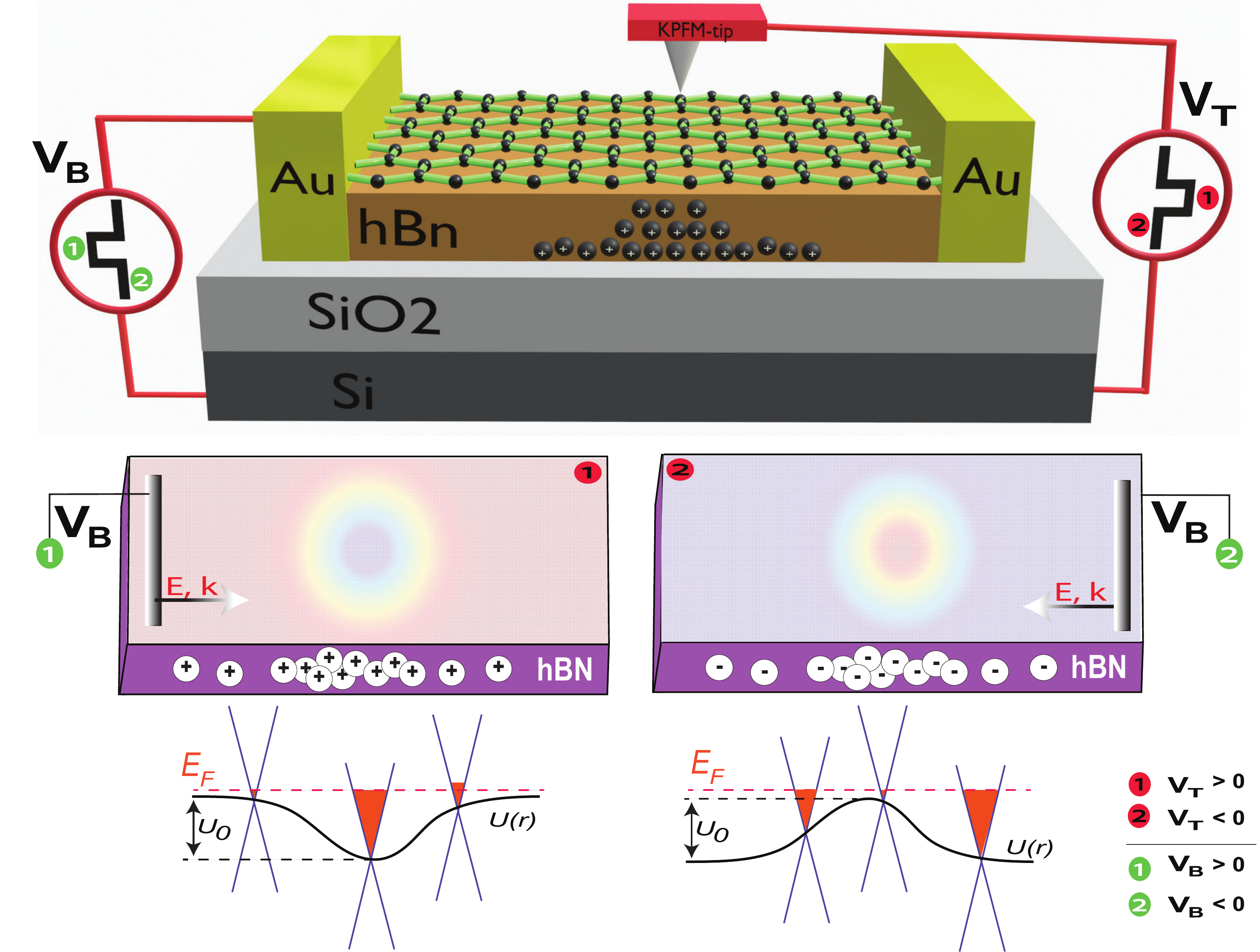}
\vspace{-0.15cm}
\caption{System device schematic. We display the device, which is a graphene sheet with zigzag boundaries that is on an hBN substrate. The back-gate substrate (source) is defined as Si, while the dielectric region is defined as SiO2. Due to the presence of a potential induced by the KPFM tip, there is a stationary distribution of charges in the hBN substrate as shown in the system.}
\label{fig1}
\end{figure}

We consider a metal-insulator-semiconductor device shown in Fig. \ref{fig1}. It is made of a graphene/hBN heterostructure deposited on an insulating substrate (SiO$_2$). The graphene charge density is contained by the type of perturbative potential. We would like to ensure a radial perturbation to trap the carriers where the perturbative potential can be generated by a tip gate electrode ($V_{T}$) using, for instance, Kelvin probe force microscopy [12,17]. The carrier density might be adjusted by tuning the tip bias making electrons or holes QBS  sensitive to the gate voltage $V_{T}$ as electrons (holes) become trapped in the hBN spacer beneath the graphene substrate under a positive (negative) bias voltage \cite{Wyatt, Li}.
%%---------------------------- Fig. 1--------------------
%%------------------------- end Fig. 1--------------------
We consider a square voltage applied with periodic modulation in the time domain, where the tip-induced potential is spread a few nanometres from the center of the graphene surface. The sample bias, denoted by  $V_B$, is used to tune the Fermi-level shift between the sample and tip gate. It is also considered to be a square voltage applied between the source (S) and drain (D) of the graphene substrate and is mainly set to be opposit to that of the tip gate electrode ($V_B \times V_{T} <0$).

The bias voltage $V_B$ is referred to by the difference in the Fermi-level between the leads (S) and (D), where we define $V_B = E_{F_D} − E_{F_S}$. In what follows, we consider $V_B=-E_{F_S}=E_F$ at the source and $E_{F_D}= 0$ at the drain.

{{\bf Tight-binding Hamiltonian }}
 We adopt a tight binding model that allows us to investigate and understand the main parameters and rules leading to valley-polarized currents depending on the values and signs of $V_B$ and $U_0$. More precisely, we show how the heterostructures can operate as valley-field effect transistors where the Fermi energy and the induced potential break the electron and hole symmetry. The Hamiltonian that describes the system, in the atomic structure of graphene and hBN layers, is given therefore, as  \cite{Marmolejo, GGiovannetti}:
%%%========== Eq.1 ===============
%\begin{equation}\label{eq1}
\begin{align}
H = & -\gamma\sum_{\langle i,j\rangle}{\bf a}_{i}^{\dagger} {\bf b}_{j} + \sum_{ \left\langle i \right\rangle }\Delta_\text{SG}\left( {\bf a}_{i}^{\dagger} {\bf a}_{i}- {\bf b}_{i}^{\dagger} {\bf b}_{i} \right)+ \nonumber \\
& \qquad \qquad \qquad \qquad \qquad \qquad \sum_{ \left\langle i \right\rangle } U_i \left( {\bf a}_{i}^{\dagger} {\bf a}_{i}+ {\bf b}_{i}^\dagger {\bf b}_{i} \right). \label{eq1}
\end{align}
%\end{equation}

where ${\bf a}_{i}$ (${\bf a}_{i}^{\dagger}$) and ${\bf b}_{i}$ (${\bf b}_{i}^{\dagger}$) are the electron annihilation (creation) operators at sublattices $A$ and $B$, respectively. Also, $\gamma$ defines the usual nearest neighbor hopping and  $U_i$ is a site dependent potential wheras $\Delta_{\rm SG}$ represents a staggered potential. The time-dependent tip voltage $V_{T}(t)$ induces a time-dependent trapped charge in the insulating hBN layer. The embedded charges beneath the graphene layer (in the hBN spacer) induced by the tip are causing a spatial variation of the charge carriers depending on its bias sign and the total screened potential $U(r,t)$ experienced by the graphene substrate is resolved self-consistently  and experimentally \cite{Orlof, Wyatt, Li}. It is used in  Eq. \ref{eq1}, and assumed to be: 
%%%========== Eq.2 ===============
\begin{eqnarray}\label{eq2}
U(r_i, \ t)\simeq U_0(t) \exp\left( -\frac{r_i^2}{R_0^2} \right)
\end{eqnarray}%{equation} 
The induced screened voltage $U_0$  in the graphene is directly controlled by the gate voltage $V_{T}$ and is spread within a diameter of radius $R_0$  (set to $55 \text{ nm}$ in our case). 

{{\bf Valley resolved current }}

In order to calculate the valley polarization, one needs to obtain the valley resolved currents $I_K$ and $I_{-K}$. To fulfill this task, we use the package kwant to obtain the scattering wavefunctions \cite{kwant} and then identify those who are of type $K$ or $K'$ at the interface of the lead (Drain). We follow these steps for each energy.
The current is therefore obtained using the following relation taking into account all contributions from different energies:

\begin{equation}\label{eq3}
I_{\pm K}  =  \frac{2q}{h}\int_{}^{} \mathcal{T}_{\pm K}(E) \times \left[ f_S\left(E, E_{F_S}\right)- f_D\left(E, E_{F_D}\right) \right]dE 
\end{equation}
where the Fermi destribution is given by,
\begin{equation}\label{eq7}
f_{i}(E, E_{F_i})=\frac{1}{1+\exp{ \left[ (E-E_{F_i})/k T\right]} },  \qquad {i\equiv S \ \text{or} \ D} 
\end{equation}
$i$ is the index of the contacts that correspond to the source (S) and drain (D) leads; $E_F$ is the carrier incident Fermi-energy; $k$ is the Boltzman constant and  $T$ temperature. The valley resolved transmission at a given energy is noted $\mathcal{T}_{\pm K}(E)$.
The valley currents are well defined for a uniform structure. IHowever, in the presence of scattering, valley mixing appear and the notion of a pure valley current can be misleading. Fortunetaly, this mixing between the valleys is in general low \cite{Jack} and can be discarded in the analysis. 
%%----------------------------------------------------------------------------------------------------
\vspace{-0.15cm}
\section{ Results and discussion}\label{sec3}
We consider the system device in Fig. \ref{fig1}, where we have a fixed width $W=110$ nm and length $L=300$ nm. The induced potential is screened by a Gaussian of radius $R_0=35$ nm \cite{Lee, Freitag, Grushevskaya}. The tight-binding parameters are set to $\gamma=-2.7 $ eV and $\Delta_\text{SG} = 29.26$ meV \cite{GGiovannetti, WYao, YRen}. Using the applied tight-binding model adopted in Eq. \ref{eq1}, we resolve the transport channels where the polarization at the Dirac cones $K_1$ and  $K_2$ are obtained independently (cf, Appendix. \ref{app-A}). 
%% -------------------------------------------------
\subsection{ Valley polarization dependence on sample bias conditions}
\label{sec3a}
\begin{figure}[tp]
\centering
%\hspace*{-0.4cm}
\includegraphics[width=8.5cm, height=4cm]{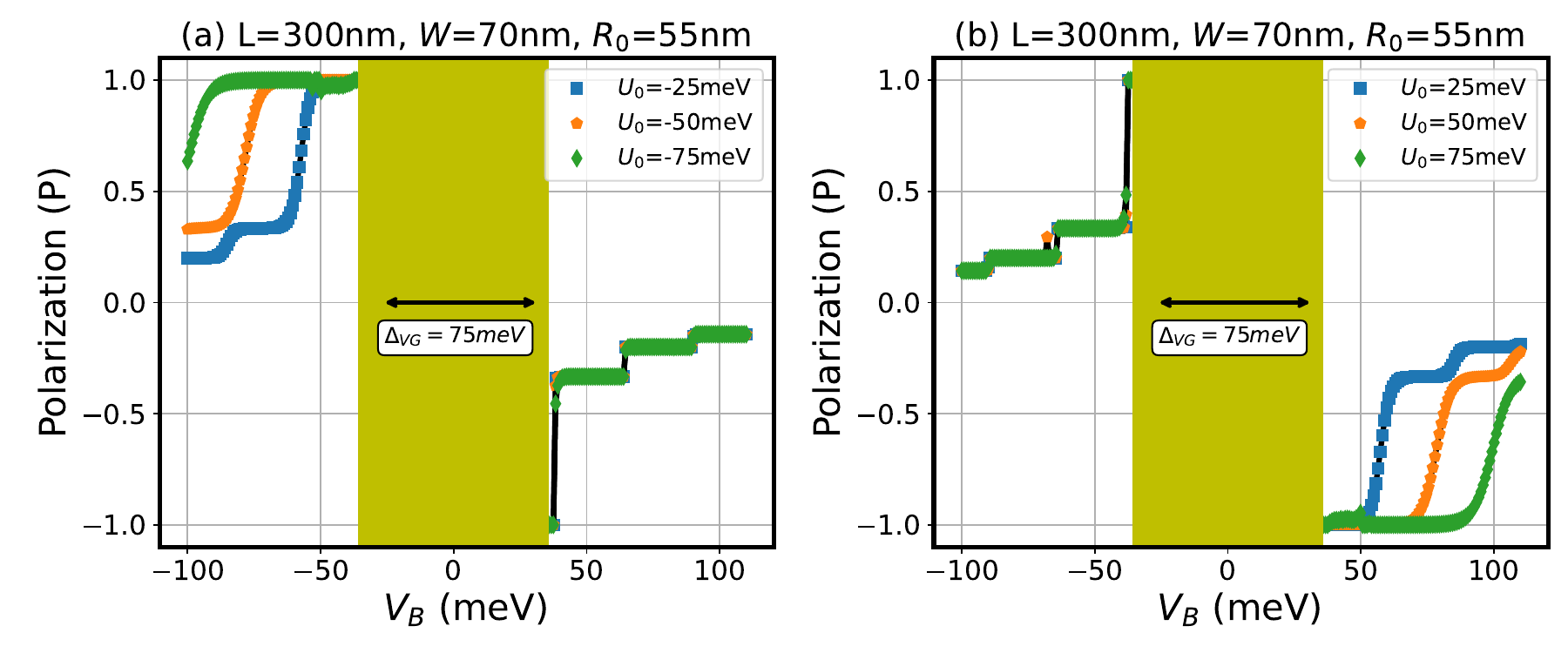}
\includegraphics[width=8.5cm, height=4cm]{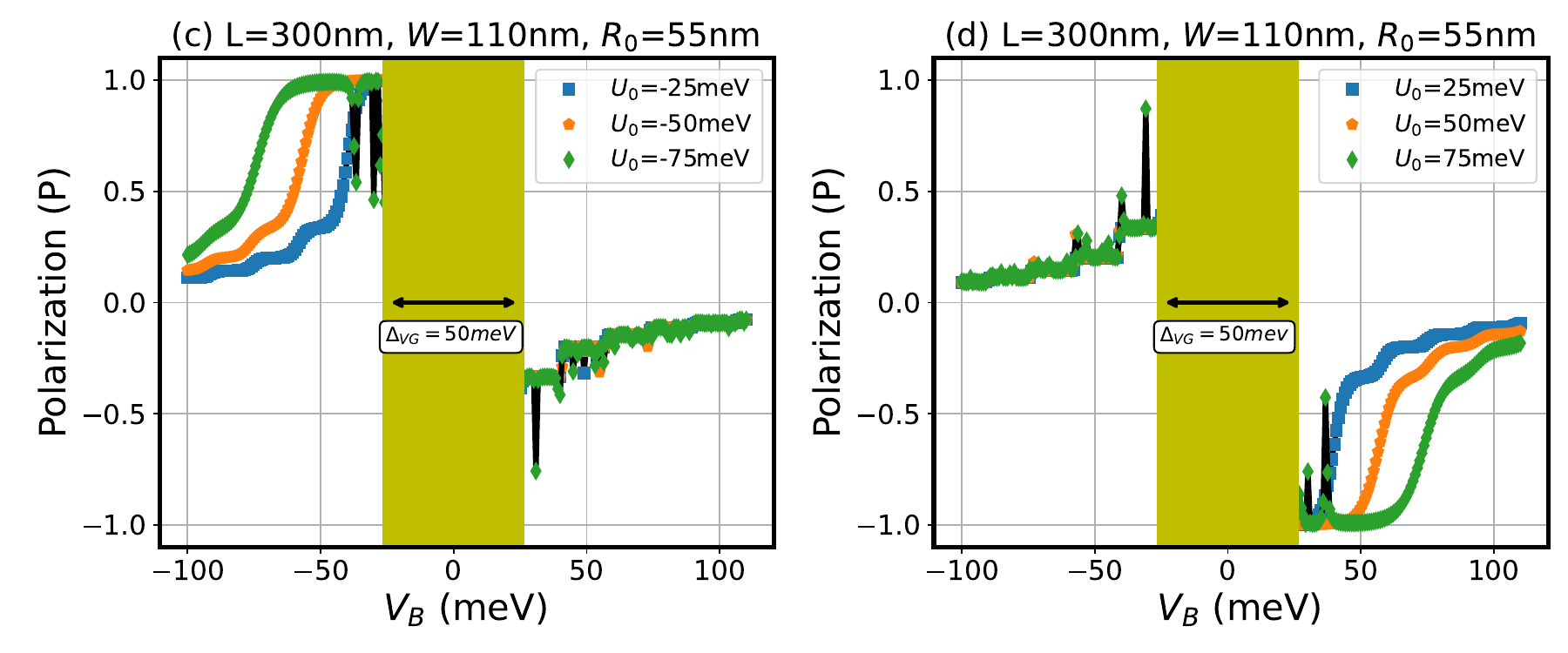}
\includegraphics[width=8.5cm, height=4cm]{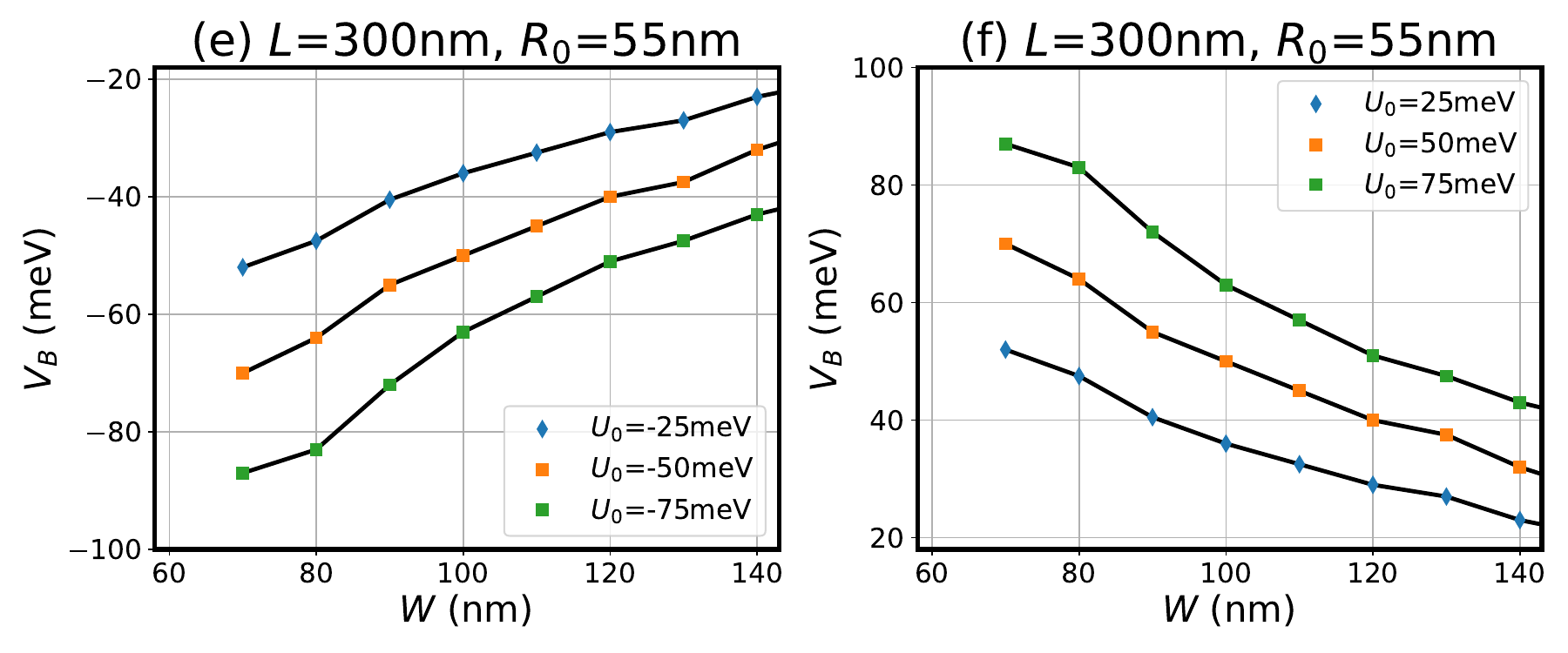}
\vspace{-0.15cm}
\caption{The figures show how the induced potential and the bias voltage monitor the valley polarization. In plots (a-d), we investigate the effect of the tip's potential $U_0$ and the bias voltage $V_B$ on the polarization of the valleys by plotting the $P(V_B)$  for several values of $U_0( \pm 25, \pm 50$, and $\pm 75)$ meV where when $P = \pm  1$ they are fully polarized. The system width is  $W = 70$ nm in (a-b) and $W = 110$ nm in (c-d) while  $R_0$ and the length of the system $L_x$ are kept fixed. The yellow regions in plots (a-d) demonstrate  that  the propagating modes contain both valleys only if Fermi-energy is beyond $|\Delta_\text{VG}/2|$. Figures (e-f) illustrate how the sample shape has a significant impact on the potential range $V^R_B$, which controls the bias range over valley polarization process. }

%it was V\R_B and I chaged it to V^R_B. Verify please
\label{fig2}
\end{figure}

Before stepping to the time-dependent response of the system, let us consider the case where we discuss the effects of strength and sign of $V_B$ and $U_0$ and observe the polarization response. It is obvious that the propagating modes, with negative velocity, sustain both valleys only if the incident Fermi-energy exceeds an energy value $E_F>  |\Delta_\text{VG}/2|$ where $\Delta_\text{VG}$ is called the valley-mode spacing gap. For this reason, the valley conductance might be computed and selected independently for $K_1$ and $K_2$ propagating channels beyond $\Delta_\text{VG}$. In this case, beyond that gap, both valley channels are allowed to scatter and hence valley polarization is straightly investigated. 
We mainly show at which conditions the induced potential and  Fermi-level monitor the valley polarization. To do so,  we set the tip bias fixed at $U_0=\pm 25, \pm 50$ or $\pm 75$ meV, and then we tune sample bias $V_B$ within a given range of few meV. 
%(2) We fix the  Fermi-energy at  $E_F=V_B=\pm 30$ or $50$ meV and we then tune $U_0$ in  a given range of few meV. 
For two width channels $W=70$ nm and $110$ nm, we compute the polarization, and the results are illustrated in Fig. \ref{fig2}. For more details about the methods utilized to output the polarization, we refer to the appendix. \ref{app-A}. 

%%---------------------------- Fig. 2--------------------
%%------------------------- end Fig. 2--------------------

According to Fig. \ref{fig2}, we observe that the valley polarization is affected by the signs of  $V_B$ and $U_0$. Indeed, the polarization is more relevant $(P=\pm 1$) if the sign of both gates is set the same; however, when the signs differ (different gate polarities), the polarization is between $-1$ and $1$; therefore, both valley channels are operating (destroyed valley filtering). Based on these statements, the valley filter is clearly operated when the product $V_{B}\times U_0$ is positive but within a bias range related to $V_{B}$, $U_0$, and sample width. 

For instance, in panels (a) and (b) of Fig. \ref{fig2}, at fixed $U_0=-75$ ($+75$) meV, the current is fully polarized  $P=+1$ ($-1$) within a potential range $−90 $ meV $<V_B< −\Delta_\text{VG}/2 $  ($ \Delta_\text{VG}/2<V_B<+90$ meV) where only $K_1$ ($K_2$) channels are conducting. Similarly, in panels (d) and (e) of Fig. \ref{fig2}, at the same fixed $U_0=-75$ ($+75$) meV, the electron are fully polarized  $P=+1$ ($-1$) but within an increased potential range $−70 $ meV $<V_B < −\Delta_\text{VG}/2$ ($\Delta_\text{VG}/2<V_B <+70$ meV). The increase in the bias range shows the effect of the sample shape on either widening or limiting the potential range used to polarize the valley current. For more details, we show in Fig. \ref{fig2} (e) and (f) how the sample shape 
%and diameter of radius $R_0$ of the screened potential 
plays an important role in increasing or lowering the potential range $V_B$ that monitors the valley polarization process. It is clearly identified that the tip gate affects proportionally the limits of the potential range whereas, the sample width effect is inversely proportional to the range of potential $V_B$.	

Based on the obtained results, it is clearly seen that the sample bias $V_B$ and tip-gate potential $U_0$ lead to operating valley polarization, where only one selected valley is allowed to pass within a given range of gates. Indeed, the range, given for a short-range potential, depends on the values and polarities of the gates as well as the shape condition related to sample and screened potential. 
To this end, if we fulfill correctly the discussed conditions, the proposed device would be able to conduct a fully valley-polarized current where only one valley is allowed, and the other one is back-scattered. 

%%---------------------------- Fig. 3--------------------
\begin{figure*}[tp]
\centering
%\hspace*{-0.4cm}
\includegraphics[width=8.9cm, height=4cm]{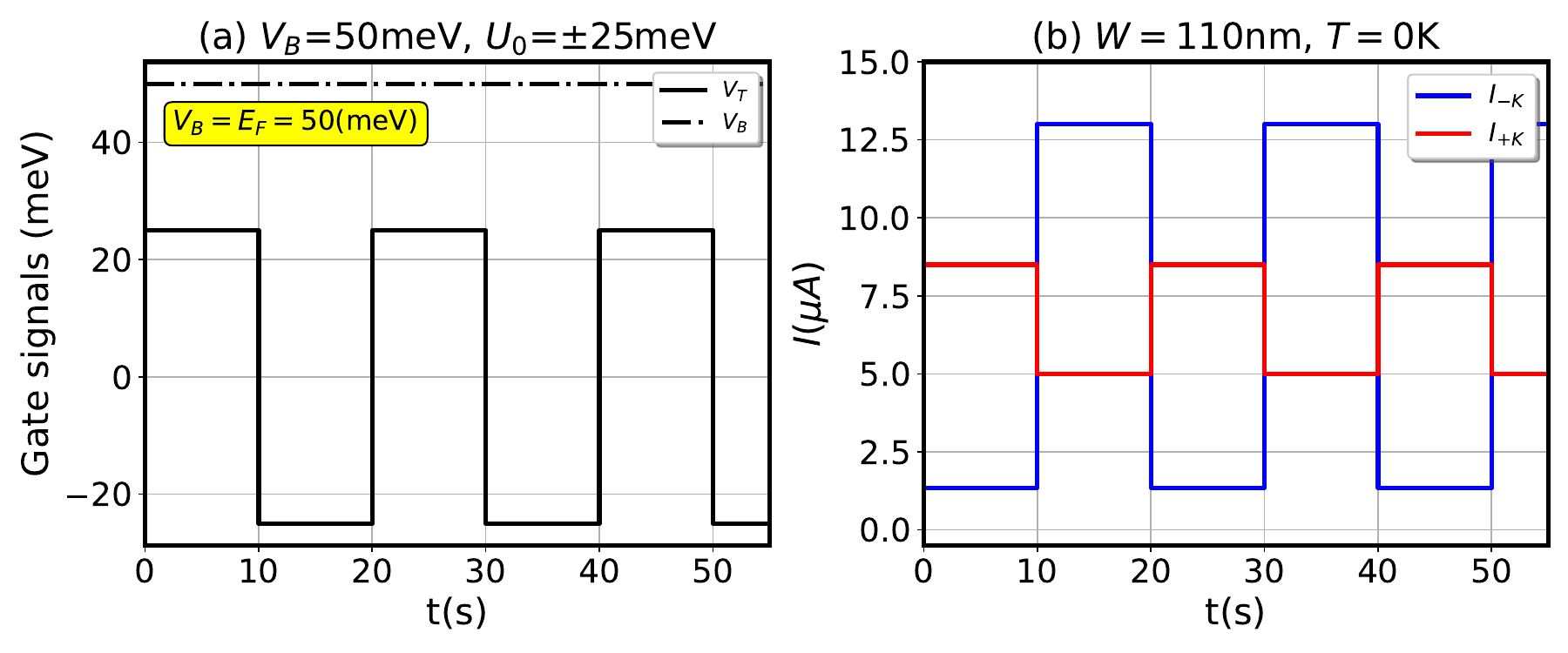}
\includegraphics[width=8.9cm, height=4cm]{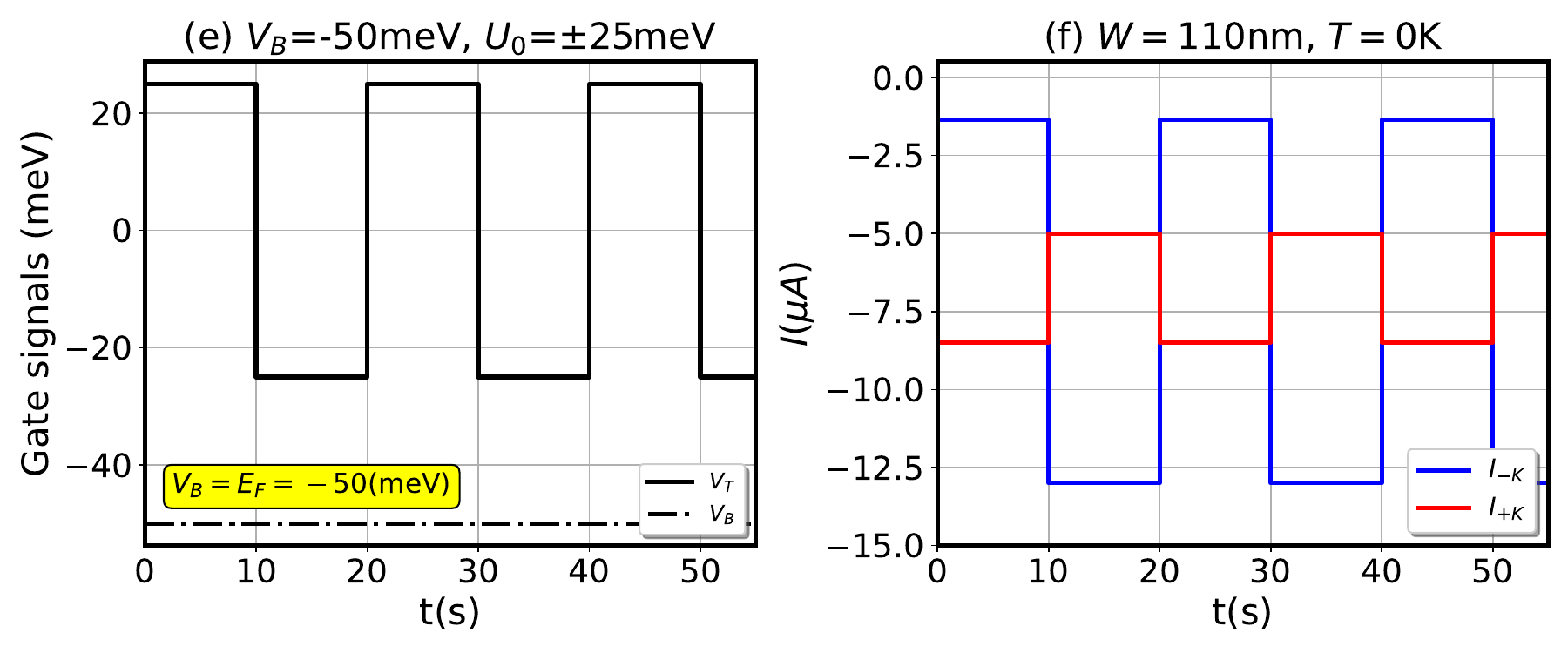}
\includegraphics[width=8.9cm, height=4cm]{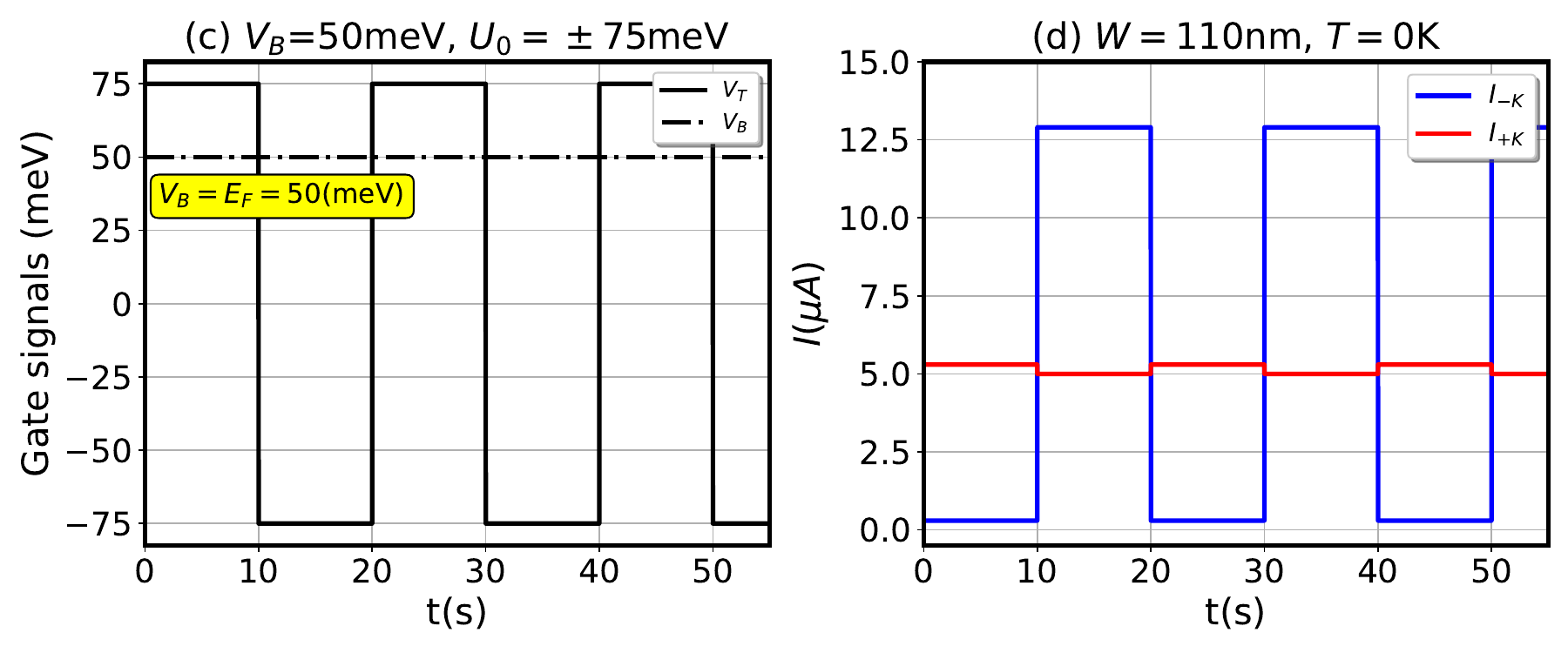}
\includegraphics[width=8.9cm, height=4cm]{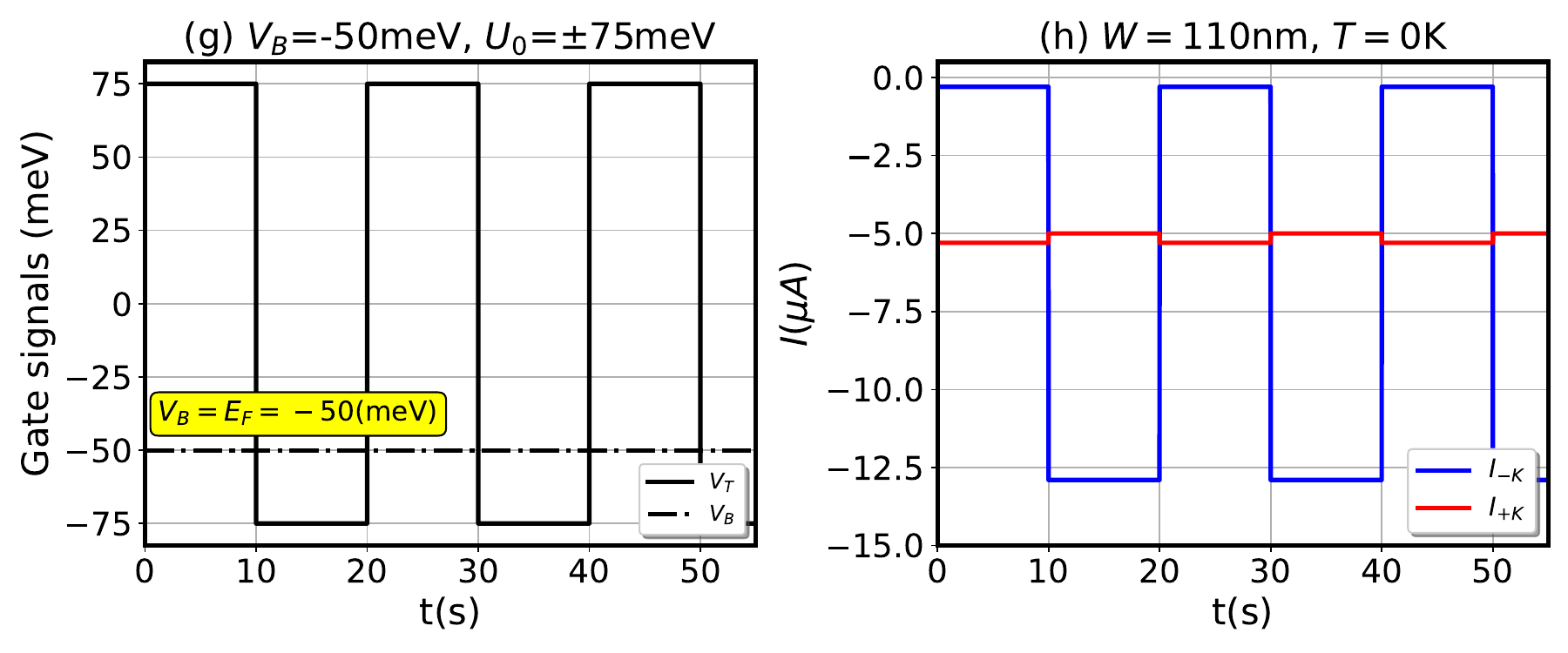}
\vspace{-0.15cm}
\caption{The valley response as the tip-potential depends on time, for various regimes. Panels (a, e) and (c, g) illustrate a periodic square tip-potential as a function of time with amplitude $U_0 = \pm 25$ meV and $U_0 = \pm 75$ meV, respectively around zero temperature and for a system's width ($W = 110$ nm), and $V_B$ are fixed ($V_B$ is equal to either $50$ meV in (a, c), or $-50$ meV in (e, g)). Plots (b, f) demonstrate the current of both valleys ($I_{+K}$ (red line) and $I_{-K}$ (blue line)) when $|V_B| > |U_0|$  while plots (d, h) show both when $|V_B| < |U_0|$. Even though the two cases behave differently, they do not provide a pure valley current.}
\label{fig3}
\end{figure*}
%%------------------------- end Fig. 3--------------------

%% -------------------------------------------------
\subsection{ Dynamic Control of valley current in several bias regimes}\label{sec3b}
Once we have discussed the conditions to operate valley transport, let us  focus in what follow on the time-dependent response. We will, particularly, focus on computing the valley current defined in Eq. \ref{eq3} for several bias regimes. 

\begin{figure*}[tp]
\centering
%\hspace*{-0.4cm}
\includegraphics[width=8.9cm, height=4cm]{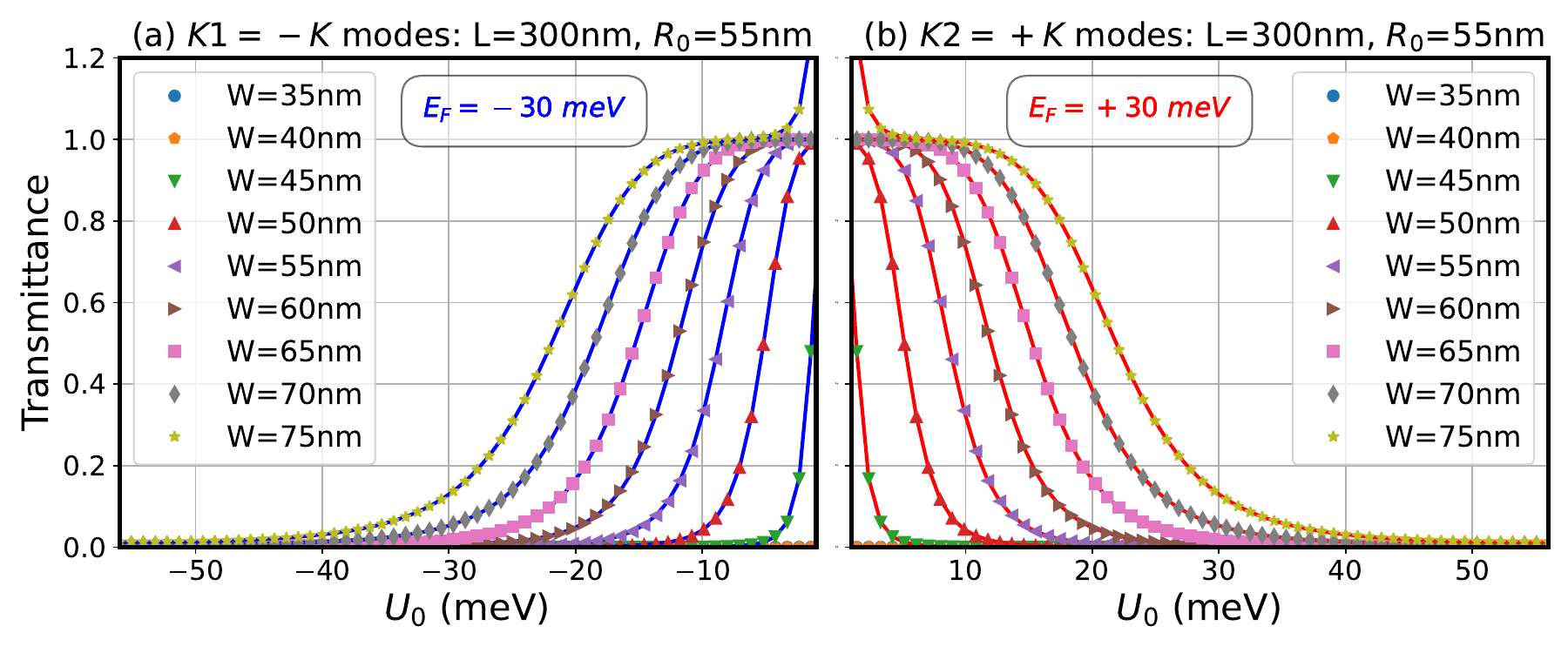}
\includegraphics[width=8.9cm, height=4cm]{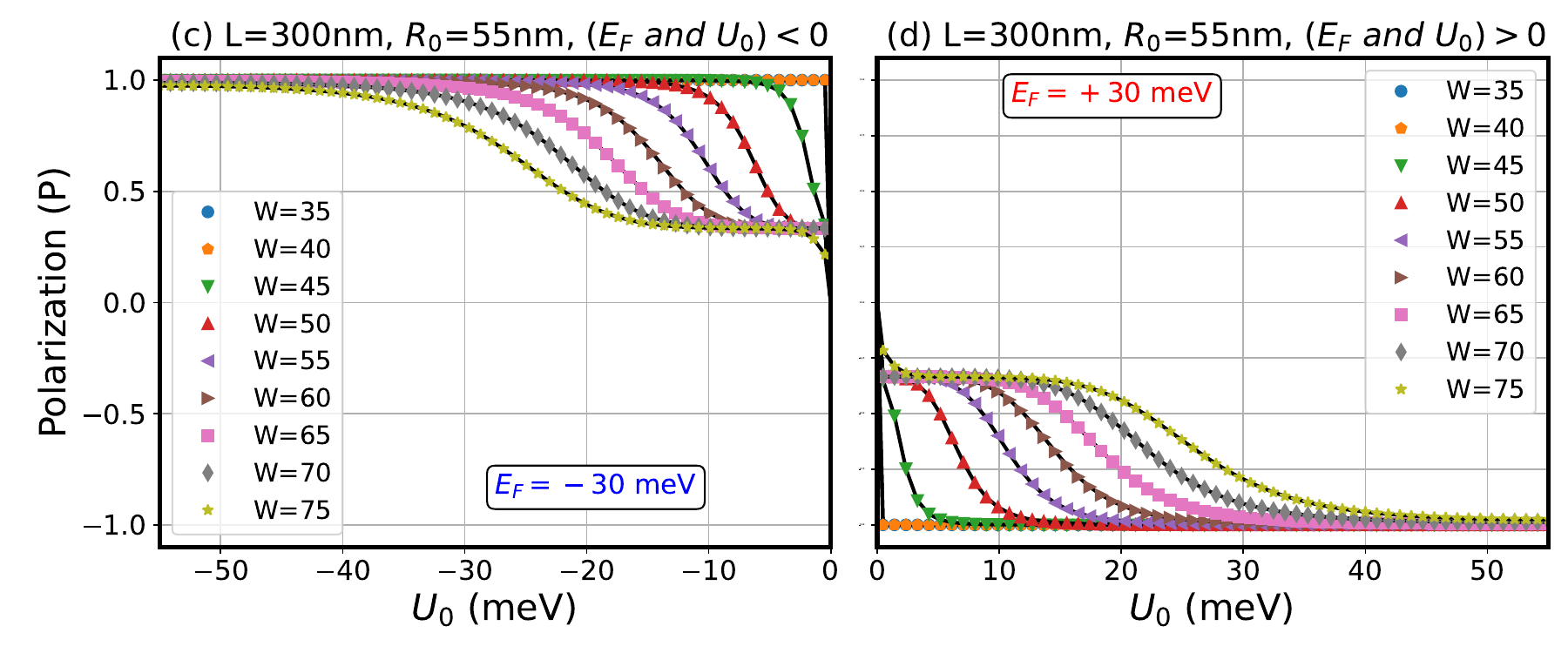}
\includegraphics[width=8.9cm, height=4cm]{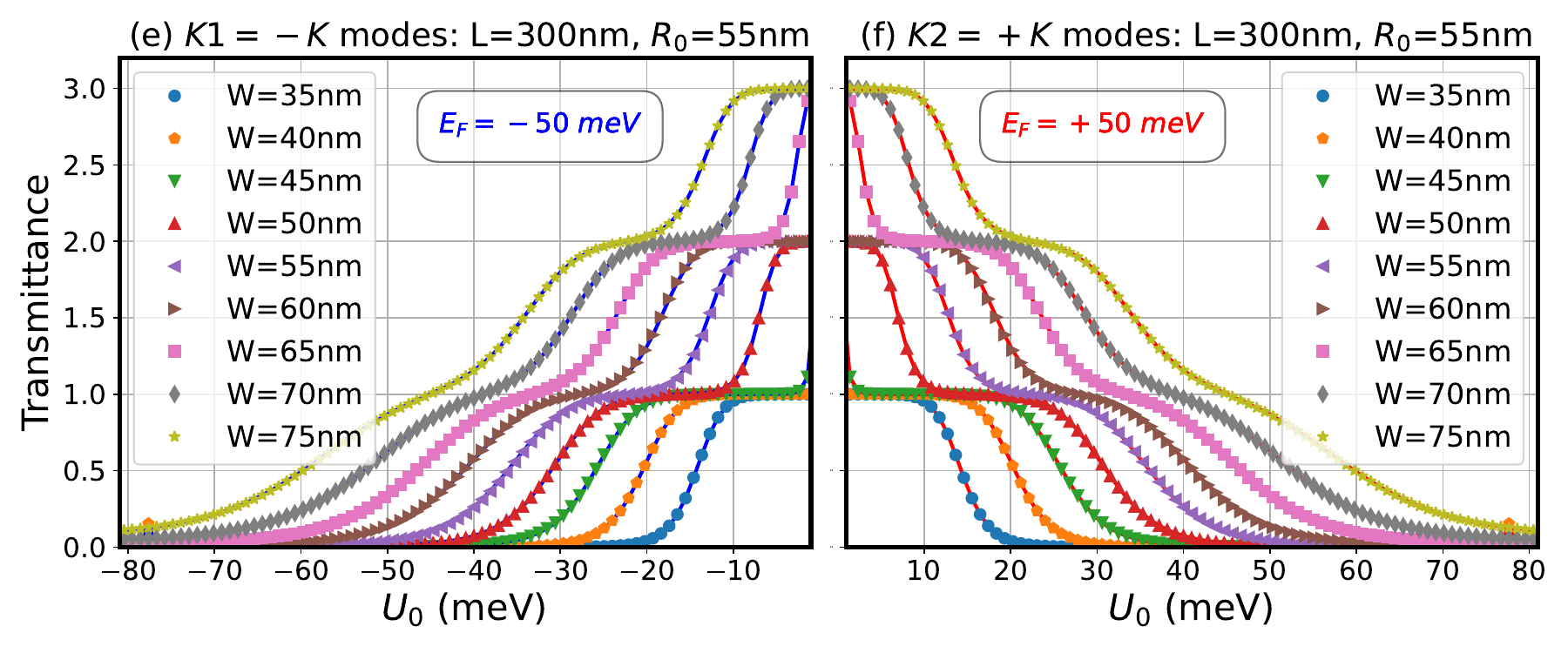}
\includegraphics[width=8.9cm, height=4cm]{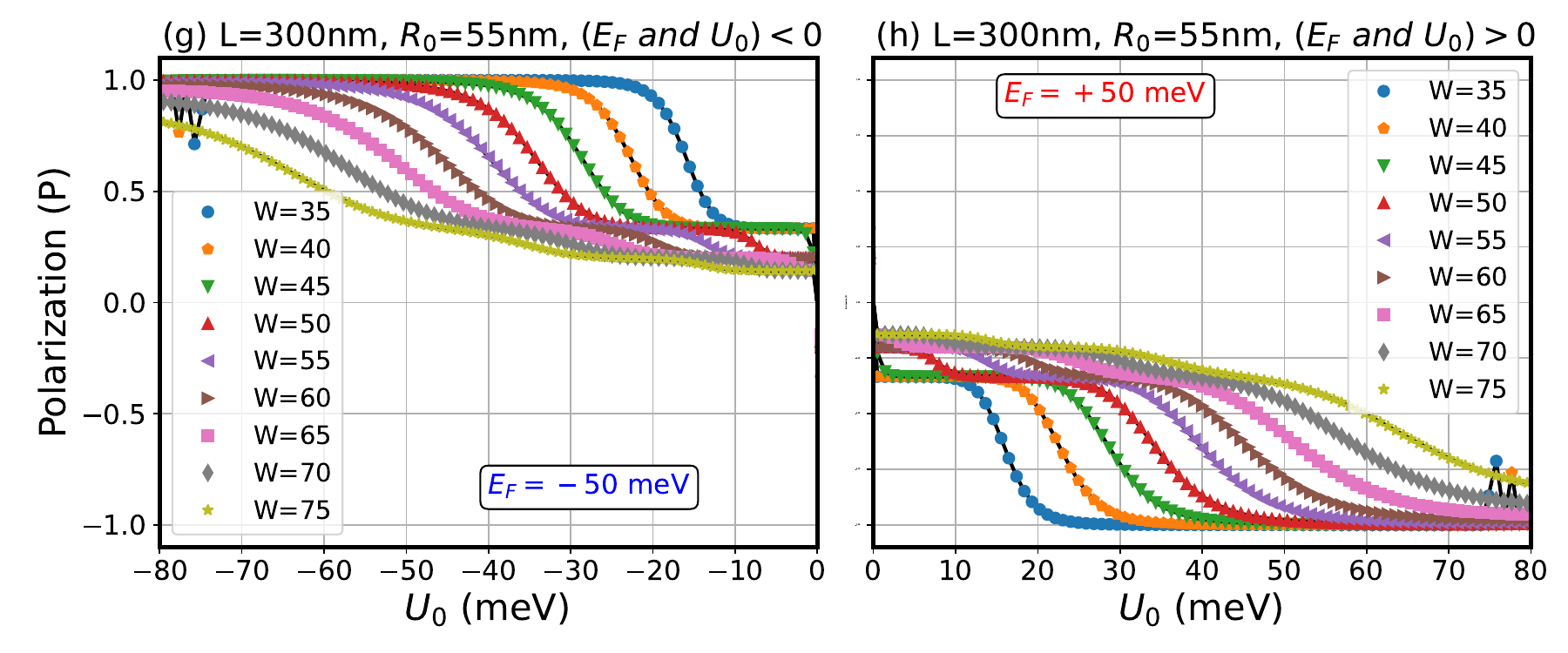}
\vspace{-0.15cm}
\caption{The impact of the system width ($W$) and the induced potential spreading $R_0$, is demonstrated by plotting the valley transmittance ({\bf Left}) and polarization ({\bf Right}) of each valley ( $-K$ and $+K$) as a function of $U_0$ for several widths. In both transmittance and polarization, we consider the sample bias $V_B = \pm 30$ meV ({\bf top panels}) and  $V_B = \pm 30$ meV ({\bf bottom panels}).}

\label{fig4}
\end{figure*}

To this end, let us discuss step by step several bias regimes and define the best conducting options in order to obtain a pure valley current in the presence of a non-zero voltage. The time-related valley response for several regimes is shown in Fig. \ref{fig3} and \ref{fig4}.

%%---------------------------- Fig. 4--------------------
%%------------------------- end Fig. 5--------------------
\subsubsection{Valley current for one bias time-dependent regime}\label{sec3b1}
The tip gate is set to output a square voltage exhibiting a periodic modulation with a time and a maximum strength (amplitude) of $U_0=\pm 25$ meV ( Fig. \ref{fig3}-a, e) or $\pm 75$ meV ( Fig. \ref{fig3}-c, g). A bias voltage is generated at the graphene surface where the sample gate is time-independent and set to a fixed bias $V_B= +50$ meV ( Fig. \ref{fig3}-a, c) or $V_B= -50$ meV ( Fig. \ref{fig3}-e, g). We assume that the period of the signal is larger than any characterestic time of the system and enough to see a steady state. This allows us to go beyond the time-dependence complications in the short transient regimes and obtain the value of the current directly from the valley-resolved conductance of the system.

The results of simulation show that for $\left| V_B \right|>\left| U_0 \right|$ ( Fig. \ref{fig3}-b, f), we wind up with charge current where both currents $I_{+K}$ and $I_{-K}$ are outputted, and hence this set of conditions does not allow to operate a pure valley current. Indeed, the propagating modes of both valleys are able to escape the screened potential barrier $U_0 = U_0 =\pm 25$ meV at $V_B = E_F = 50$ meV. However, by increasing the screened potential barrier to a higher value $U_0=U_0=\pm 75$ meV ( Fig. \ref{fig3}-d, h), the propagating modes of both valleys at $E_E=V_B=\pm50$ meV, will be processed differently, due to electron-hole broken symmetry, (see Sec. \ref{sec3} for more discussion) and only one valley is allowed to escape to the screened potential barriers while the other one is back-scattered. This happens only when both biases are polarized with the same sign. Inverting the sample bias winds up with inverting the valley polarization and current direction. 

Clearly seen that the bias options shown in Fig. \ref{fig3} do not provide pure valley current. They do only lead to either $I_{+K}$ or $I_{-K}$ current in the modulation at $(2n+1)T/2<t<(2n)T/2$, and it outputs charge current in the modulation at $(2n+1)T/2<t<(2n)T/2$ where $n$ is an integer, and $T$ is the signal period ($T=20$s). Inverting the sample bias does invert the valley polarization, but it always outputs a charge current at the second modulation as depicted in Fig. \ref{fig3} (b) and (f). The type and direction of the valley are defined and controlled by the sample bias where tip-gate is used to either operate ($V_{B}\times U_0>0$) or destroy ($V_{B}\times U_0<0$) the valley filter as shown in \ref{fig3} (d) and (h). Based on these remarks, a pure valley current might be generated only if we set the bias $V_B=0$, in Fig. \ref{fig3} (c) and (g),  at time modulation $t=2n(T/2)$ but unfortunately, this option is not a good choice in electronic applications. 

Based on the discussed regimes in Fig.\ref{fig3}, a pure valley current is only allowed if: \\
(1) the sample bias is bigger than the tip bias ( $\left| V_B \right|>\left| U_0 \right|$) \\
(2) both gates are defined as the negative or positive of the gate (same polarity) \\
(3) the gates are put to zero bias at the second time modulation, where this latter is not a good option in digital electronics.

Broadly speaking, the generation of the valley-current, as shown in Fig.\ref{fig3}, is not only ensured by the values of $V_{B}$ and $U_0$. In fact, the sample width and the spreading of the induced potential controlled by $R_0$ both play a predominant factor in allowing or destroying the valley process. To clearly show this point, we compute the valley conductance in terms of $U_0$ for several ribbon widths and set the induced potential width $R_0=55$ nm and then fix the sample bias at (1) $V_B=\pm 30$ meV as shown in Fig.\ref{fig4} (a-d), or (2) $V_B=\pm 50$ meV as depicted in Fig.\ref{fig4} (h-d). Clearly seen, that the valley conductance is depending on the sample width within a given energy range that depends on the sample set as depicted in Fig.\ref{fig4} and previously discussed in Fig. \ref{fig2} (c) and (f). 

First, according to Fig.\ref{fig4} (a) and (b), at $E_F=\pm 30$ meV, the conductance is either one or zero (in the unit of $2 e^2/h$). Additionally, the polarization is controlled and might be set to either one or less than one, depending on the tip gate and sample width, as shown in Fig.\ref{fig4} (c) and (d). More precisely, it is noticed that for sample widths $W<R_0=55$ nm, at $E_F=\pm 30$ meV, the system provides fully polarized current if we set the tip gate to $\left|U_0\right| \leq \pm 50$ meV, since $P(W<R_0, \ \left|U_0\right|=\pm 50 \text{ meV})=\pm 1$. Therefore, the system exhibits a smooth valley dependence which confirms the existence of valley current in Fig. \ref{fig3}. However, when the tip gate is set to $\left|U_0\right|>\pm 50$ meV, the system does not seem to work effectively for $W<R_0=55$ nm where the overall polarization drops to zero $-1<P\left ( W<R_0, \ \left|U_0 \right|>\pm 50 \text{ meV} \right )<1$ and therefore no important valley-selective process will be noticed. 

Second, at $E_F=\pm50 $ meV the valley conductance steps from $3$ to $2$ then from $1$ to $0$, in the unit of $2e^2/h$ (Fig. \ref{fig4} (e) and (f)). A better valley modulation is given by the blue, orange, green, red, and purple dotted lines ($35$ nm$<W<55$ nm) where the polarization in Fig. \ref{fig4} (g) and (h) shows a fully polarized conductance where we obtain $P=\pm 1$ for the tip bias $abs(U_0)>23$ meV ($W=35$ nm), $abs(V{T})>30$ meV ($W=40$ nm), $abs(U_0)>38$ meV ($W=45$ nm), $abs(U_0)>46$ meV ($W=50$ nm), $abs(U_0)>55$ meV ($W=50$ nm), respectively.

Based on these discussions, we clearly confirm that the system is able to operate a pure-valley current once particular conditions related to lattice shape, induced potential width (spreading), and range of the voltage gates are correctly chosen. Furthermore, this presentation of various data sets would help experiments to set the best parameters that allow constructing a valley-built device that outputs a pure valley current.

Fortunately, we can select a variety of those conditions where the pure-valley current might be generated. However, the key point leading to having different {\bf ON/OFF} valley currents shifted in time is still not discussed and it is mainly fulfilled if the pulse frequency and both voltage values demonstrate a valley transistor functionality. This point will be discussed, in the following subsection, where both gates are chosen to output  synchronized nonzero voltage over time.
%%---------------------------- Fig. 5--------------------
\begin{figure}[tp]
\centering
%\hspace*{-0.4cm}
\includegraphics[width=9cm, height=4.5cm]{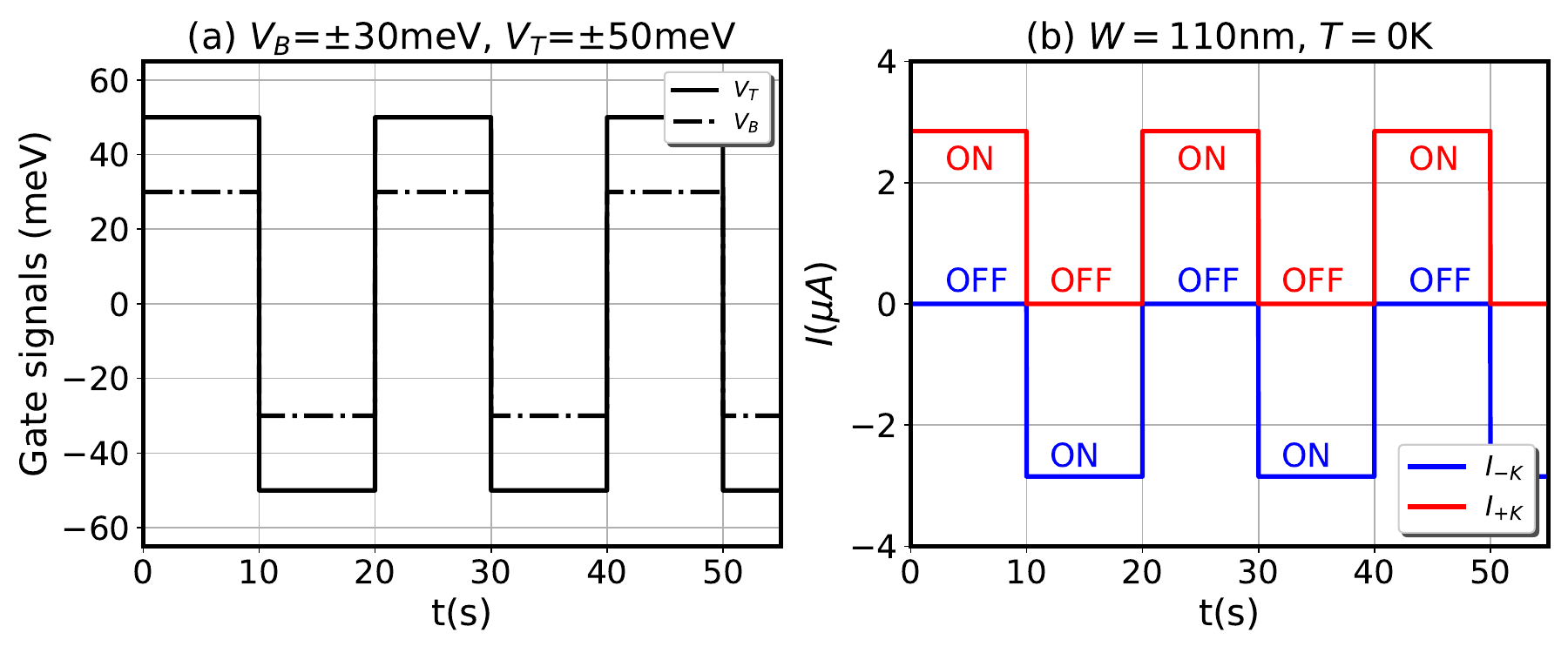}
\includegraphics[width=9cm, height=4.5cm]{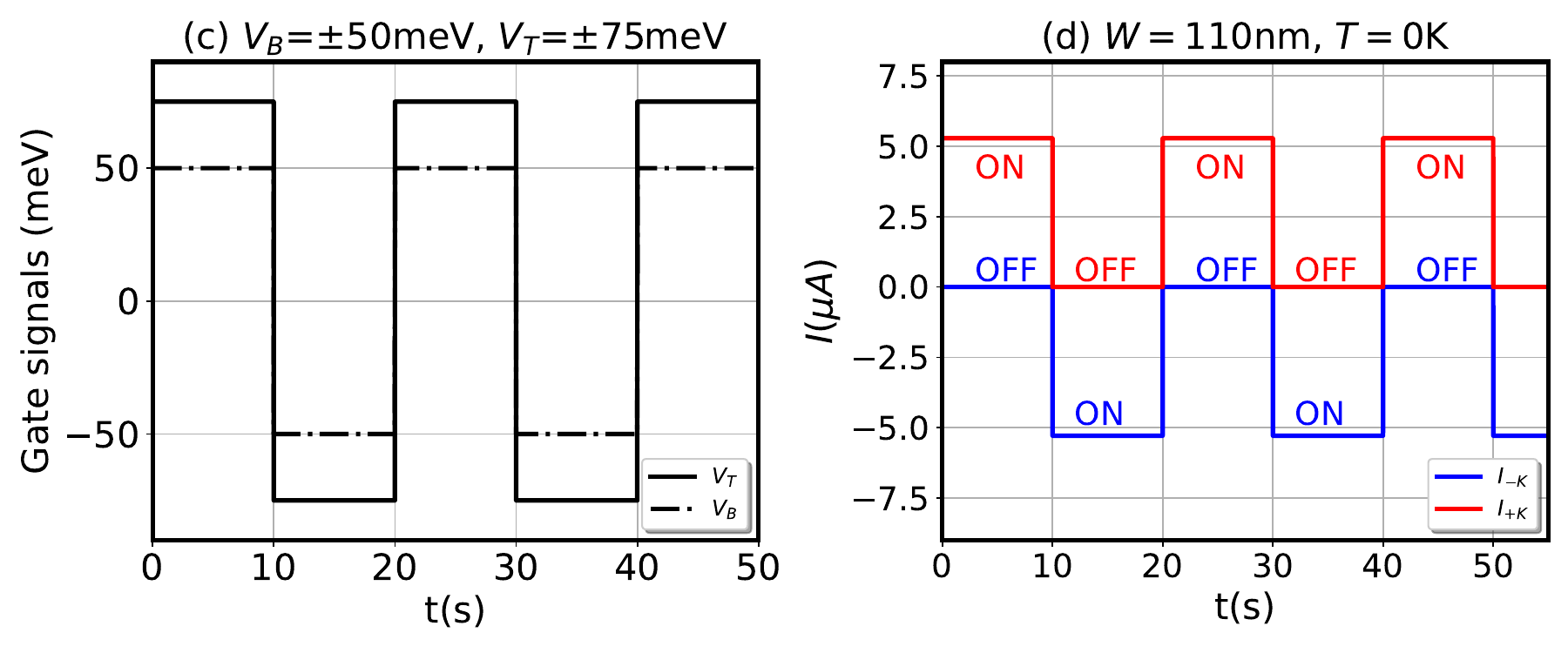}
\vspace{-0.15cm}
\caption{Plots (a, c)  illustrate a periodic square tip-potential as a function of time with amplitudes $\left\{ U_0 = \pm 50; \ V_B = \pm 30 \right\}$ meV and $\left\{ U_0 = \pm 75; \ V_B = \pm 50 \right\}$  meV, respectively, where in these cases the gates are set to the same polarity and synchronized with the same frequency. Plots (b, d) show the correspondence current of both valleys ($I_{-K}$ and $I_{+K}$), where  plot (b) represents case (a)  while plot (d) is for case (c). At nonzero voltages, the device responds with various {\bf ON/OFF} pure valley currents as the gates synchronize with a periodic modulation in time.}
\label{fig5}
\end{figure}
%%------------------------- end Fig. 5--------------------
\subsubsection{ Dynamic control of valley current in nonzero time-dependent regimes}\label{sec3b2}

Finally, we demonstrate that a pure valley current can be generated even when both gate biases are nonzero. This is achieved by applying time-dependent voltages to both gates, ensuring that the tip and bias potentials maintain the same sign through synchronized signal frequencies, that is, identical periods and polarities in their amplitudes. To implement this, we consider a square voltage signal  that undergoes periodic time modulation while remaining fixed at nonzero values throughout the cycle
with (1) $V_B=\pm 30$ meV and $U_0=\pm 50$ meV (Fig. \ref{fig5}-a)  or (2)  $V_B=\pm 50$ meV and $U_0=\pm 75$ meV (Fig. \ref{fig5}-c). For both cases in (1) and (2), the time-related valley current is shown in Fig. \ref{fig5} at zero temperature.%$T=10K$.

It is clearly demonstrated that synchronizing the sample and tip voltage with the same frequency and polarity induced prominent characteristics of pure valley-polarized current even in nonzero bias where either $I_{+K}$ or $I_{-K}$ are present over time but with the different flow direction. More precisely, as the gates output synchronized voltages with a periodic modulation in time, the device responds in different {\bf ON/OFF} valley currents at nonzero voltages accordingly. Interestingly, when the $I_{-K}$ is turned ON (forward-current) the $I_{+K}$ is turned OFF. Inverting the polarity of the voltages flips the states of the currents: $I_{-K}$ is turned OFF, and $I_{+K}$ is turned ON (reversed-current).

One can observe that the strength of the valley current can be enhanced by increasing either the bias range or the amplitude of the applied square potentials. For instance, (1) at $V_B=30$ meV, the device outputs a valley current of strength $I=2.8\ \mu A$, as shown in panel (b) of Fig. \ref{fig5}; (2) at $V_B=50$ meV, the device outputs a valley current $I=5.2 \ \mu A$, as shown in Fig. \ref{fig5} (d).

The current characteristics, in Fig. \ref{fig5}, show that the adopted MIS type device, using graphene/h-BN heterostructure, exhibits the behavior of a valley-transistor where the {\bf ON/OFF} valley-current is controlled by both gates and operates independently and inversely only within some conditions related to the device gates polarity, voltage range, shape condition, and range of the induced potential. Notably, the gate modulated of the valley current is observed at low gate biases of few meV, which makes it a good device to operate at low power and leads to building a valley circuit for low power operation. 	

%%---------------------------- Fig. 6--------------------
\begin{figure*}[tp]
\centering
%\hspace*{-0.4cm}
\includegraphics[width=17cm, height=8cm]{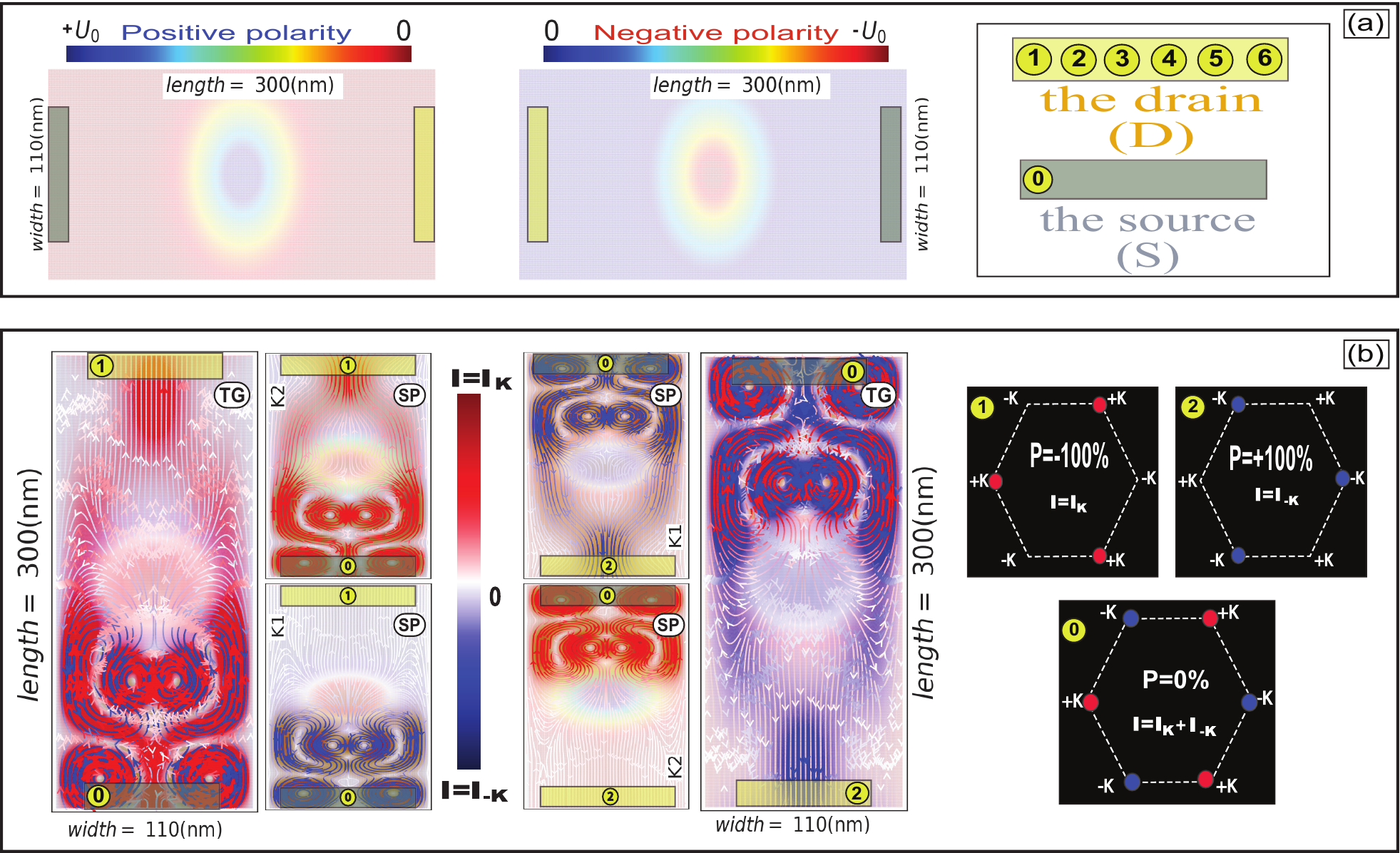}
\vspace{0.25cm}
\includegraphics[width=17cm, height=5.5cm]{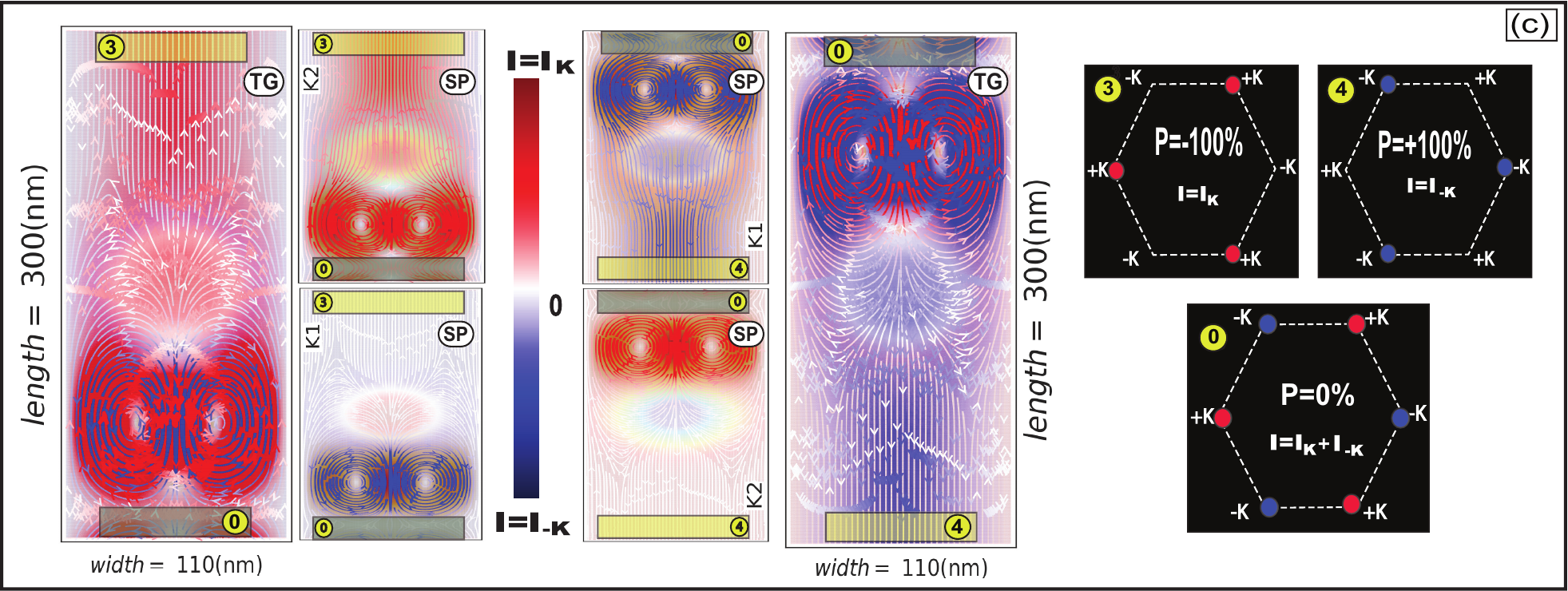}
\vspace{0.25cm}
\includegraphics[width=17cm, height=5.5cm]{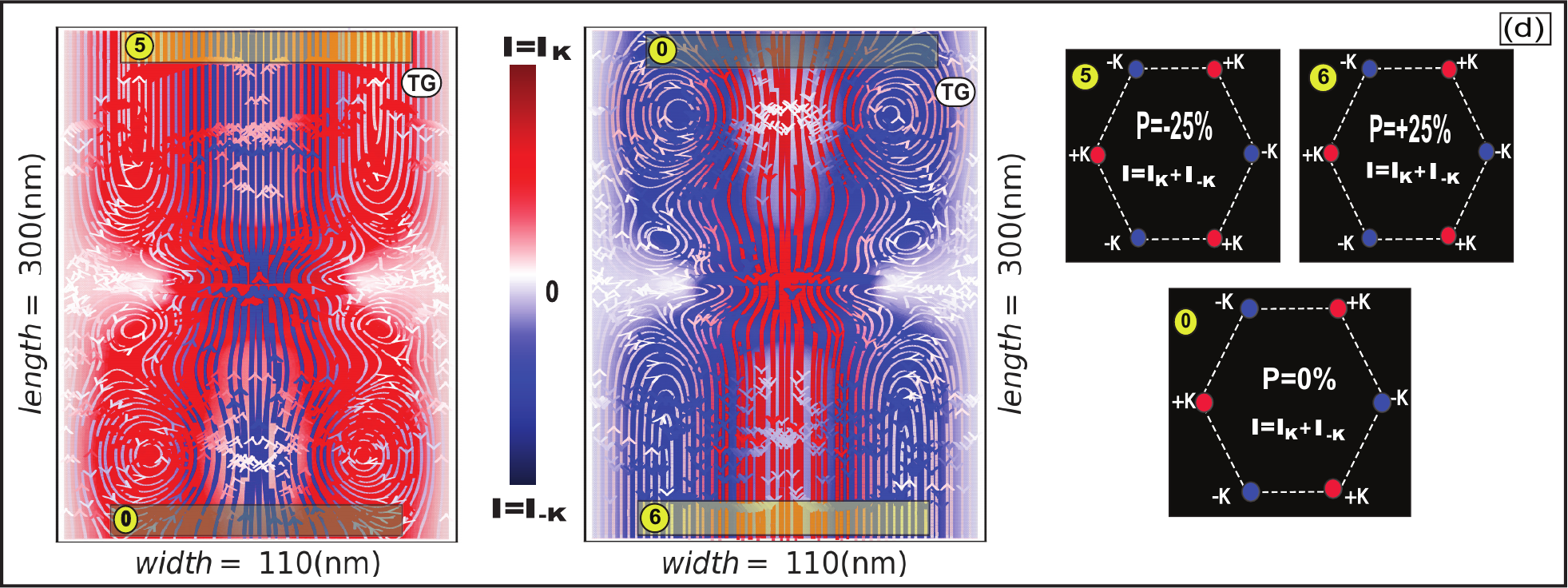}
\vspace{-0.15cm}
	\caption{
		In panel (a) we map the induced potential $U(r)$ raised from the stationary distribution of the charges in the hBN substrate, due to the KPFM tip where the mapping is for the positive and negative polarity of $U_0$. The gray and yellow shaded regions define the source and drain and they are also used to provide the projection defined in Eq. \ref{eqC4} (cf, Appendix. \ref{app-C}).	
		In panels (b-d) we define the real-space mapping of both current in {\bf Left} and {\bf Middle}  where the blue (red) line show the $K_1 = −K$ ($K_2 = +K$) current. {\bf TG} refers that we plot both currents; however, {\bf SP} means that we plot separately each valley current for better visualization. The spectral valley polarization is shown in the  {\bf Right} panels. 	
		Panel (b) is for the set $\left\{ U_0 = +70; \ V_B = +50 \right\}$ meV in {\bf Left} and $\left\{ U_0 = -70; \ V_B = -50 \right\}$ meV in {\bf Middle}. Panel (c) for the set $\left\{ U_0 = +50; \ V_B = +30 \right\}$ meV in {\bf Left} and $\left\{ U_0 = -50; \ V_B = -30 \right\}$ meV in {\bf Middle}. Panel (d) is for the case where the tip gate is set to the negative of the bias gate, given by the set $\left\{ U_0 = -50; \ V_B = +30 \right\}$ meV in {\bf Left} and $\left\{ U_0 = +50; \ V_B = -30 \right\}$ meV in {\bf Middle}. }\label{fig6}
\end{figure*}
%%---------------------------- Fig. 6--------------------

\subsection{ Real-space mapping of the valley current and spectral valley polarization}\label{sec3c}
To further endorse the presence of pure-valley current in the proposed device, we show in more detail how the valley current and polarization around the Dirac point are obtained. Also, we plot the real space map related valley current and polarization, at a given time, for the discussed regimes in section. \ref{sec3b}. For more details, about the method used for real space mapping of the spectral valley polarization and the valley current, we kindly refer to the appendix. \ref{app-C} and \ref{app-D}.

We compute, in Fig. \ref{fig6} % and Fig. \ref{fig8}  t
the valley current and polarization %at the positions highlighted by %square boxes 
related to the regimes in Fig. \ref{fig5} (b, d) and Fig. \ref{fig4} (b, f). We calculate the spectral density at each valley, and then show the Fourier maps for the sites in the scattering region closer to the lead (D). 

For the regimes depicted in Fig.~\ref{fig5}, the valley filtering always occurs when both signals are gated with either positive $\{ V_B=50\text{~meV} \ \text{or} \ 70\text{~meV}; \ U_0=30\text{~meV} \ \text{or} \ 50\text{~meV} \}$ or negative sign $\{ V_B=-50\text{~meV} \ \text{or} \ -70\text{~meV};\ U_0=-30\text{~meV} \ \text{or} \ -50\text{~meV}\}$.

In this case, only one valley is allowed to pass through the Gaussian potential, as shown in the current maps in panels (b) and (c) of Fig. \ref{fig6}. In these cases, the transmitted modes that belong to the electrons in the $K_1=-K$ ($K_1=+K$) valley are clearly shown in their corresponding polarization mapping and controlled by the bias polarity $V_B<0$ ($V_B>0$). 

More precisely, the valley filtering is fully operating for the potential set given as $\left\{ V_B=\pm 50 \ \text{meV},  U_0=\pm 70 \ \text{meV} \right\}$, shown in Fig. \ref{fig6} (b) and $\left\{ V_B=\pm 30 \ \text{meV},  U_0=\pm 50 \ \text{meV} \right\}$, depicted in Fig. \ref{fig6} (c). For both cases, it is clearly observed that, at $(2n)T/2<t<(2n+1)T/2$ in Fig. \ref{fig5} (a, c), only the $K_2=+K$ valley current crosses the tip-induced potential in the forward direction to reach the drain while the other valley current is backscattered toward the source. Conversly, at $(2n+1)T/2<t<(2n)T/2$ only $K_1=-K$ valley current is mapping the sample but passing in the backward direction. These results validate our previous assumptions, which state that the bias regimes illustrated in Fig. \ref{fig5} define the best conducting options to obtain a pure valley current in the presence of a non-zero voltage, and therefore a valley transistor behavior takes place.

From the other side, the valley filtering is not perfectly operating for the bias regimes given in opposite polarities, where in this case, it is clearly observed, at  $(2n)(T/2)<t<(2n+1)(T/2)$  in Fig. \ref{fig4}, both valleys are allowed to pass through the Gaussian potential, as shown in the mapped current of Fig. \ref{fig6} (d). We have $I_K$ and $I_{-K}$ current, which are mapping the sample and crossing the tip-induced potential with (1) a polarization $P=(-25\%)$ for the set $\left\{ V_B=+30 \ \text{meV},  U_0=-50 \ \text{meV} \right\}$ and (2) a polarization $P=(+25\%)$ for the set $\left\{ V_B=-30 \ \text{meV},  U_0=+50 \ \text{meV}  \right\}$. Therefore, the regime in which the bias gate has a polarity opposite to that of the tip gate  is not suitable for effective operation of a valley transistor device.

\section{ Conclusion and summary }\label{sec4}

The results presented in this study demonstrate the functionality of a valleytronic device based on a graphene/hBN two-dimensional heterostructure. We conducted a systematic investigation of valley transport in this setup and showed that it could serve as a fundamental building block for valley field-effect transistors (VFETs), capable of operating under low bias conditions while enabling controllable {\bf ON/OFF} switching of valley-polarized currents.

We provided a detailed interpretation of the mechanisms underlying this behavior, highlighting the critical role of parameters such as the bias range, polarity, and frequency of the tip and sample gate voltages, as well as the sample width and the spatial extent of the tip-induced potential landscape in graphene/hBN. When these parameters are properly tuned, the device can efficiently generate a periodic, valley-polarized current with opposite valley indices at alternating time intervals, even under finite bias conditions. Notably, this is achieved with very low power consumption, on the order of a few meV.

We primarily investigated the conditions under which the induced potential and the Fermi level govern the valley polarization. Specifically, we considered the valley polarization parameter $P\in \left[-1, \ +1 \right]$
where $P=+1$ corresponds to a scattered electron from the $-K$ valley and $P=-1$ to one from teh $+K$ valley
Furthermore, we presented real-space maps of the current distribution and valley polarization curves, illustrating how valley currents are either transmitted or blocked as they encounter the perturbative tip-induced potential landscape.

Importantly, the spectral valley polarization reveals that the $-K$ valley current fully dominates the region defined by the perturbative potential landscape induced by Kelvin probe force microscopy. In this configuration, the $K_2 = +K$ valley current is entirely deflected away from the perturbed region, resulting in 100\% of the current being carried by the $-K$ valley: $I = I_{-K}$. Remarkably, reversing the polarity of the gate voltages reverses the transport behavior. In this case, the spectral valley polarization shows that the $+K$ valley current becomes dominant within the potential landscape, while the $-K$ valley current is completely excluded, yielding $I = I_{+K}$.

\medskip

In summary, this study demonstrates the feasibility of integrating valley-functionality into graphene/hBN-based two-dimensional heterostructures. We successfully achieved valley-selective transport, enabling the device to switch between an ON-state for the $-K$ valley and OFF-state for the $+K$ valley, and vice versa, solely by tuning the gate polarity.

\section{Acknowledgments}
The authors acknowledge computing time on the supercomputer SHAHEEN at KAUST Supercomputing Centre and the team assistance. A. A. and A. B. gratefully acknowledge the support provided by the Deanship of Research
Oversight and Coordination (DROC) at King Fahd University of Petroleum and Minerals(KFUPM) for funding this work through the Intelligent secure system center research grant No. INSS2512. J.L. acknowledges the support of the Natural Sciences and Engineering Research Council of Canada (NSERC) RGPIN-2022-03882 and (NRC) AQC-200-1.

%%----------------------------------------------------------------------------------------------------
%%%%%%%%%%%%%%%%%%%%%%%%%%%%%%%%%%%%%%%%%%
\vspace{6pt}
%%%%%%%%%%%%%%%%%%%%%%%%%%%%%%%%%%%%%%%%%%
%%%%%%%%%%%%%%%%%%%%%%%%%%%%%%%%%%%%%%%%%%

\clearpage
\appendix
\numberwithin{equation}{section}
\begin{widetext}
\begin{appendices}
		%======================================================================================
		\section{The non-equilibrium Green’s function method for the current flow}\label{app-A}
		The flow of current in graphene sheets is investigated by using the non-equilibrium Green’s function (NEGF) technique, which is based on the tight-binding Hamiltonian (Eq. \ref{eq1}) and the hopping parameter of the graphene. 
		In this appendix part, we give a short overview of the relevant parts of NEGF method [1a,2a, 3a] and adjust it for resolving the valley related transport channels.
		The Green's function of the system is given by
		\begin{equation}\label{eqA1}
		G(K)=\frac{1}{E-H(K)-\Sigma_R - \Sigma_L}
		\end{equation}
		where $E$ is the energy of the coming electrons, $H$ is Hamiltonian Eq. \ref {eq1}, $\Sigma_R$ ($\Sigma_L$) is the self-energy of the right lead (left lead) and the self-energies describe the effect of the attached leads on the central system. 
		
		The wave functions of the propagating states $\phi(K)$ are obtained directly by exploiting the Kwant simulation software \cite{kwant},  which allow us to assess the valley transport features by exploring various propagation modes based on their momentum space and connect them with the scattering region and then go to the other edge of the nonribbon as shown in Fig. \ref {fig1}.  For these contacts we use the wide-band model with
		\begin{equation}\label{eqA2}
		\Sigma_i=-\mathrm{i} \sum_{n \in C_{\mathrm{i}}}|n\rangle\langle n|\  \quad \text{and} \quad \Gamma^{\mathrm{i}}=-2 \mathfrak{Im}(\Sigma_{\mathrm{i}})  
		\end{equation}
		Where the index $\mathrm{i}$ might be either the for the left (L) lead or the right (R) lead. The transmission (or conductance) between the source lead and the lead at the other edge is then given by
		\begin{equation}\label{eqA3}
		T(E) =\operatorname{Tr}\left(G \Gamma^{L} G^{\dagger} \Gamma^{R}\right)   
		\end{equation}
		The flowing local current through the system is calculated by
		\begin{equation}\label{eqA4}
		\text{I}_{ij}=\mathfrak{Im}(t_{ij}^*G_{ij}^{n})
		\end{equation}
		where
		\begin{equation}\label{eqA5}
		G^{n}= G \Gamma^S G^{\dagger}
		\end{equation}
		$G^n$ is the quantum electron density and $S$ represents the source lead. 
		
		Since, we seek to  apply a tight-binding model by resolving valley transport channels in the reciprocal space, we will mainly compute separately the electron transmission probability at each Dirac cone ($K_1 = −K$ and $K_2 = +K$). To do so, we consider the states characterized by its valley  where we  select the propagating modes  (1) around $K_1$ obtained from the wave functions of the propagating states  $\phi({\bf k} < 0, v < 0)$ and (2) around $K_2$ obtained from the wave functions of the propagating states  $\phi({\bf k} > 0, v < 0)$. Therefore, the electron transmission probability from left to right or (right to left)  might be observed separately within the Green’s function approach and the valley resolved channels lead to the electron's valley transmittance $T_{-K}$ (for $K_1$ channels) and $T_{+K})$ (for $K_2$ channels):
		\begin{equation}\label{eqA6}
		T_{\pm K}^{m, n}= \operatorname{Tr}[G^m_{\pm K} \Gamma G^n_{\pm K} \Gamma^T  ], \qquad (m, n = L, R)
		\end{equation}
		After getting the propagating modes, the two valleys can be separated depending on the momentum sign and therefore the suitable representation for the valley polarization is obtained as
		\begin{equation}\label{eqA7}
		P = \frac{T_{-K} -T_{+K} }{T_{-K} + T_{+K}}
		\end{equation}
		For $P = \pm1$, full polarized transmittance is guaranteed since the electrons are completely confined at the $\pm K$ valleys, while for $P = 0$, the transmitted electrons are completely unpolarized.

		\section{Mapping of the valley polarization in real space}\label{app-C}
		Mapping the valley polarization is of great interest since it allows to observe the presence or absence of pure-valley current in the real space map of the given system. This latter is computed by projecting the  propagating states of  system wave  functions  $\phi(K)$ onto the eigenstate of the graphene lattice $\psi({\bf k})$. In this case, one can determine the valley polarization $P({\bf k})$ of a such state $\phi({\bf k})$ as 
		\begin{equation}\label{eqC1}
		P({\bf k}) = \left| \left\langle \psi({\bf k})|\phi({\bf k}) \right\rangle \right|^2
		\end{equation}
		where the graphene eigenstate states are defined analytically as:
		\begin{equation}\label{eqC2}
		\psi_n({\bf k})= \begin{cases}e^{i {\bf k} r_n} & n \in A \\ \beta e^{i{\bf k}\left(r_n-\delta\right)} \frac{g({\bf k})}{|g({\bf k})|} & n \in B\end{cases}  
		\end{equation}
		A and B are the sublattices of graphene, $\beta$ is the sign of E and $\delta=(-a_{cc}/2, \sqrt{3}a_{cc}/2) $ is the vector that connects the two sublattices A and B where $a_{cc}$ is graphene nearest distance. Finally, g({\bf k}) is given by
		\begin{equation}\label{eqC3}
		g({\bf k})=-t e^{-3 i k_x a_{cc}}\left[1+2 e^{3 i k_x a_{cc} / 2} \cos \left(\frac{\sqrt{3} K_y a_{cc}}{2}\right)\right]\;\;\;\;
		\end{equation}
		We can identify the valley polarization thanks to the representation of the occupied states in k-space provided by the projection 
		$P_i({\bf k})$ that can be written as
		\begin{equation}\label{eqC4}
		P({\bf k}) = \left\langle \psi^{\dagger} | G \Gamma^S G^{\dagger}  | \psi \right\rangle_{[[R]]} 
		\end{equation}
		Using this equation and implementing the NEGF formalism, we can obtain the projection over  a certain region $[[R]]$ of the system.  
		%=================================================================
		\section{Mapping of the valley current in real space}\label{app-D}
		To map the valley current in real space, we calculate the local current by collecting the valley states separately using the procedure adopted in Kwant package \cite{kwant}. To this end, we define the valley current $J_{\pm K}^{ab}$ as 
		
		\begin{equation}\label{eqD1}
		J_{\pm K}^{ab}=\Phi^{*}({{\bf k_{\binom{ > 0}{ < 0}}, v<0}})\left(i \sum_{\gamma}^{}H^{*h}_{ab\gamma} H^h_{a\gamma}-H^h_{a\gamma}H_{ab\gamma} \right) \Phi({{\bf k_{\binom{ > 0}{ < 0}}>0, v<0}}),
		\end{equation}
		\noindent where $H_{ab}$ is a matrix with zero elements except for those connecting the sites a and b. In this case, the hopping matrices in the heterostructure are obtained from the first term of Eq. \ref{eq2}. 
		The valley current $J^{a, b}$ in Eq. \ref{eqD1} has been derived from the continuity equation using the density operator $\rho_q$  and expressed as
		\begin{equation}\label{eqD2}
		\rho_q({\bf k}<0, \ \text{or} \ >0)=\sum_{a}\Phi^{*}_{a}({\bf k}<0, \ \text{or} \ >0) H^h_{q} \ \Phi_{a}({\bf k}<0, \ \text{or} \ >0)=0,
		\end{equation}
		and
		\begin{equation}\label{eqD3}
		\frac{\partial \rho_q({\bf k}<0, \ \text{or} \ >0)}{\partial t}-\sum_{b}^{}J^{a, b}({\bf k}<0, \ \text{or} \ >0)=0.
		\end{equation}
		where $H^h$ is the Hamiltonian of the heterostructure in the scattering region whose size is $N_1 \times N_2$ sites and $\Phi(v<0)$ is the wave eigenstates of the propagating modes of the graphene’s lead whose size is $N_1$. Here $q$ defines all sites or hoppings in the scattering region.  $H_{ab}$ is a matrix with zero elements except for those connecting the sites a and b. In this case, the hopping matrices in the heterostructure are obtained from the first term of Eq. \ref{eq2}.
\end{appendices}

%\clearpage

%{\bf References}
\end{widetext}
\bibliography{references}

\end{document}